\documentclass[aps,twocolumn,superscriptaddress,floatfix,prc]{revtex4-2}
\usepackage[utf8]{inputenc}
\usepackage[pdfencoding=auto,psdextra]{hyperref}
\usepackage{amsmath}
\usepackage{amssymb}
\usepackage{mathtools}
\usepackage{physics}
\usepackage{graphicx}
\usepackage{url}
\usepackage{blindtext}
\usepackage{units}
\usepackage[dvipsnames]{xcolor}
\usepackage{pifont}
\usepackage{simplewick}
\usepackage{wasysym}
\usepackage{placeins}
\usepackage{comment}
\usepackage{xifthen}
\usepackage{anyfontsize}
\newcommand{\cre}[2]{{\hat{#1}}^\dagger_{ #2}}
\newcommand{\anh}[2]{{\hat{#1}}^{\;}_{ #2}}
\newcommand{\NO}[1]{\left\{#1\right\}}
\newcommand{\sket}[1]{\ket{#1}}
\newcommand{\sbra}[1]{\bra{#1}}
\newcommand{\sfrac}[2]{\frac{#1}{#2}}
\newcommand{\expect}[1]{\langle #1 \rangle}

\DeclareMathOperator*{\argmin}{arg\,min}

  \def\nuc#1#2{\relax\ifmmode{}^{#1}{\protect\text{#2}}\else${}^{#1}$#2\fi}
  \def\itnuc#1#2{\setbox\@tempboxa=\hbox{\scriptsize\it #1}
    \def\@tempa{{}^{\box\@tempboxa}\!\protect\text{\it #2}}\relax
    \ifmmode \@tempa \else $\@tempa$\fi}

\newcommand{\MeV}{\,\unit{MeV}}
\newcommand{\hw}{\ensuremath{\hbar\Omega}}
\newcommand{\Nmax}{\ensuremath{N_\mathrm{max}}}
\newcommand{\NNLOsat}{N2LO$_\mathrm{sat}$}

\newcommand{\newabbreviation}[3]{\newcounter{#1}\expandafter\newcommand\csname#1\endcsname[1][]{\ifthenelse{\equal{##1}{abreviate}}{#2}{\ifthenelse{\isempty{##1} \AND \value{#1}=0 \OR \equal{##1}{explain}}{#3 (#2)\stepcounter{#1}}{\ifthenelse{\equal{##1}{fullname}}{#3}{#2}}}}}

\newabbreviation{NOTB}{\ensuremath{\mathrm{NO2B}}}{normal-ordering two-body}
\newabbreviation{SRNOTB}{SR-NO2B}{single-reference normal-ordering two-body}
\newabbreviation{MRNOTB}{SR-NO2B}{multi-reference normal-ordering two-body}
\newabbreviation{HO}{HO}{harmonic oscillator}
\newabbreviation{NCSM}{\text{NCSM}}{no-core shell model}
\newabbreviation{CI}{CI}{configuration-interaction}
\newabbreviation{CC}{\text{CC}}{coupled-cluster}
\newabbreviation{NNLO}{N2LO}{next-to-next-to-leading order}
\newabbreviation{VNO}{VNO}{vacuum-normal-ordered}
\newabbreviation{CoM}{\ensuremath{\mathrm{CM}}}{center-of-mass}
\newabbreviation{threeNF}{\text{3NF}}{three-nucleon force}
\newabbreviation{twoNF}{\text{2NF}}{two-nucleon force}
\newabbreviation{chiEFT}{\(\chi\)EFT}{chiral EFT}
\newabbreviation{NZME}{NZME}{non-zero matrix elements}
\newabbreviation{SD}{\text{SD}}{Slater determinant}

\newcommand{\xiCM}{\ensuremath{\xi_{\CoM}}}
\newcommand{\wxi}{\ensuremath{\omega_{\xi}}}
\newcommand{\NCM}{\ensuremath{N_{\CoM}}}
\newcommand{\wN}{\ensuremath{\omega_N}}
\newcommand{\wtrap}{\ensuremath{\omega_{\rm{trap}}}}

\begin{document}

%-------------
\title{Normal-ordering approximations and translational (non) invariance}

%\thanks{This manuscript has been authored by UT-Battelle, LLC under
%  Contract No. DE-AC05-00OR22725 with the U.S. Department of
%  Energy. The United States Government retains and the publisher, by
%  accepting the article for publication, acknowledges that the United
%  States Government retains a non-exclusive, paid-up, irrevocable,
%  world-wide license to publish or reproduce the published form of
%  this manuscript, or allow others to do so, for United States
%  Government purposes. The Department of Energy will provide public
%  access to these results of federally sponsored research in
%  accordance with the DOE Public Access
%  Plan. (http://energy.gov/downloads/doe-public-access-plan).}

% The authors may supply their own affiliations
\author{T. Dj\"arv}
\author{A. Ekstr\"om}
\author{C. Forss\'en}
\affiliation{Department of Physics, Chalmers University of Technology,
  SE-412 96 Gothenburg, Sweden}
\author{G. R. Jansen}
\affiliation{National Center for Computational Sciences, Oak Ridge
  National Laboratory, Oak Ridge, Tennessee 37831, USA}
\affiliation{Physics Division, Oak Ridge National Laboratory, Oak Ridge, TN 37831, USA}
%-------------
\begin{abstract}
  Normal-ordering provides an approach to approximate three-body
  forces as effective two-body operators and it is therefore an important
  tool in many-body calculations with realistic nuclear interactions.
The corresponding neglect of certain three-body terms in the
normal-ordered Hamiltonian is known to influence translational
invariance, although the magnitude of this effect has not yet been
systematically quantified.
In this work we study in particular the normal-ordering two-body
approximation applied to a single harmonic-oscillator reference state.
We explicate the breaking of translational invariance and demonstrate
the magnitude of the approximation error as a function of model space
parameters for \nuc{4}{He} and \nuc{16}{O} by performing full no-core
shell-model calculations with and without three-nucleon forces.
We combine two different diagnostics to better monitor the breaking of
translational invariance. While the center-of-mass effect is shown to
become potentially very large for \nuc{4}{He}, it is also shown to be
much smaller for \nuc{16}{O} although full convergence is not
reached. These tools can be easily implemented in studies using other
many-body frameworks and bases.
\end{abstract}

\maketitle

%-------------
\section{Introduction}
The need for an effective \threeNF{} to describe the strong nuclear interaction in
atomic nuclei is well established~\cite{hammer2013}. Its origin dates
back to Fujita and Miyazawa, who computed the \threeNF{} arising from
a two-pion exchange diagram~\cite{fujita1957}.
In the modern understanding, \threeNF{}s arise in effective field
theories (EFTs) as a consequence of integrating out degrees of
freedom. More specifically, \threeNF{}s appear in the \chiEFT{} of the
strong nuclear interaction at \NNLO{} and above in standard power
counting schemes of the chiral
expansion \cite{epelbaum2006,epelbaum2009,machleidt2011,hammer2019}.
Besides the EFT arguments, it has also been shown that several
experimental findings are difficult to reproduce without the inclusion
of a \threeNF{}, such as certain three-nucleon scattering
observables~\cite{kalantar2012}, the $A=3,4$ binding
energies~\cite{tjon1975,Wesolowski:2021cni}, and selected light
nucleus spectroscopy~\cite{navratil2003,barrett2013,carlson2015}.

Unfortunately, the full inclusion of \threeNF{}s in quantum
many-body methods is computationally demanding due to the large
increase in the number of \NZME{}~\cite{vary2009}. In fact, the
escalation of memory demands and the increase in execution time often
render solutions intractable when explicit \threeNF{}s are added.
This situation has initiated searches for approximation schemes that
will include the most important physics of \threeNF{}s, but at a lower
computational cost.

One such approximation scheme is the \SRNOTB{}
approximation~\cite{hagen2007a, roth2012}, which potentially can
incorporate the dominant piece of the \threeNF{} as an effective
\twoNF{} and therefore at significantly reduced computational cost.
This is often done by approximating the ground state with a single
\SD{} and then normal ordering the \threeNF{} relative to this
reference state using Wick's theorem~\cite{wick1950}. The expectation value of
the residual three-nucleon term, acting only outside the references state,
is assumed to give a much smaller contribution to the ground-state
energy than the induced two-, one- and zero-body parts---and is
therefore discarded.

The \SRNOTB{} approximation has been used with great success in
\emph{ab initio} nuclear structure calculations---in particular to
facilitate studies of medium-mass
systems~\cite{roth2012,binder2013b,hagen2014,jansen2014,hergert2016}.
Normal-ordering approximations beyond the single reference state have
also been developed~\cite{cipollone2013, gebrerufael2015}.
The accuracy of the \SRNOTB{} has been benchmarked, e.g., in
Refs.~\cite{hagen2007a,roth2012,binder2013}. The importance of
residual \threeNF{}s was shown to be small by explicit comparison with
calculations using full \threeNF{}s. However, these benchmarks were
performed at a fixed oscillator frequency and the dependence on
model-space parameters has not been investigated. This is
  particularly important since we show that the sensitivity of
  \SRNOTB{} to the choice of basis frequency could be significant.
We stress that our concern pertains to the explicit breaking of
translational invariance of the underlying Hamiltonian due the \NOTB{}
approximation.  The ensuing \CoM{} dependence is therefore of
different origin compared to the well-known problem of \CoM{} mixing
as a consequence of e.g. particular truncations of the single-particle
basis that are used in some many-body
solvers~\cite{gloeckner1974,hagen2009a, roth2009b, hergert2016,
  parzuchowski2017, tichai2018, hoppe2020}.

In this work, we have studied the \SRNOTB{} approximation in a \HO{}
basis with a \NCSM{} total-energy truncation. We consider the
closed-shell systems \nuc{4}{He} and \nuc{16}{O}---for which the
single-reference approximation is appropriate---and we explore the
accuracy of the \NOTB{} approximation and the breaking of translational
invariance as a function of model space parameters \Nmax{} and \hw.
The realistic \NNLOsat{} interaction~\cite{ekstrom2015a} with both
\twoNF{} and \threeNF{} terms is used for
all numerical calculations unless otherwise stated.

The full Hamiltonian, the \NCSM{} method, and the \SRNOTB{} approximation is
introduced in Sec.~\ref{sec:Theory}. The \CoM{} problem is
presented in Sec.~\ref{sec:the-cm-problem} where we also introduce and
benchmark the metrics that will be used in the analysis. The \NOTB{}
approximation errors for ground-state energies and radii for
\nuc{4}{He} and \nuc{16}{O} are analyzed in Sec.~\ref{sec:results},
while concluding remarks are given in Sec.~\ref{sec:discussion}.

%-------------
\section{Theory} \label{sec:Theory}
\subsection{The Hamiltonian}
The general Hamiltonian that is considered in this work can be written

\begin{equation}
  \hat{H} = \hat{T}_{\rm{int}} + \hat{V}_{\rm{\twoNF}} +
  \hat{V}_{\rm{\threeNF}}.
  \label{eq:hamiltonian}
\end{equation}
The potential operators are expressed in second quantized form as
\begin{equation} \label{eq:2nf_vac_no}
\hat{V}_{\rm{\twoNF}} = \sfrac{1}{4}\sum_{\substack{\alpha\beta \\ \alpha'\beta'}}
\sbra{\alpha\beta}{V}_{\rm{\twoNF}}\sket{\alpha'\beta'}
\cre{c}{\alpha}\cre{c}{\beta}\anh{c}{\beta'}\anh{c}{\alpha'},
\end{equation}
and
\begin{equation} \label{eq:3nf_vac_no}
\hat{V}_{\rm{\threeNF}} =
\sfrac{1}{36}\sum_{\substack{\alpha\beta\gamma \\ \alpha'\beta'\gamma'}}
\sbra{\alpha\beta\gamma}{V}_{\rm{\threeNF}}\sket{\alpha'\beta'\gamma'}
\cre{c}{\alpha}\cre{c}{\beta}\cre{c}{\gamma}
\anh{c}{\gamma'}\anh{c}{\beta'}\anh{c}{\alpha'},
\end{equation}
with Greek letters representing tuples of the well-known
single-particle quantum numbers $(n,l,j,j_z,t_z)$ in a \HO{}
basis. These operators, as well as the intrinsic kinetic energy,
$\hat{T}_{\rm{int}}$, depend on relative coordinates (in position and
momentum space) such that the Hamiltonian~\eqref{eq:hamiltonian} is
translationally invariant.
\subsection{The no-core shell model}

To solve the many-body Schrödinger equation we employ the \NCSM{} in
which the Schrödinger equation
\begin{equation}
  \hat{H}\ket{\Psi} = E\ket{\Psi},
\end{equation}
is rewritten as a finite matrix eigenvalue problem by expanding the
eigenstates of the Hamiltonian $\hat{H}$ in a finite many-body basis
\(\{\ket{\Phi_i}\}_{i=1}^D\), i.e., the \NCSM{} eigenstate $n$ is
\begin{equation}
  \ket{\Psi_n}_{\NCSM} = \sum_{i = 0}^D c_{n,i}\ket{\Phi_i}.
  \label{eq:ncsm_eigenstate}
\end{equation}
The \SD{} many-body basis state \(\ket{\Phi_i}\) is constructed using
second-quantization
\begin{equation}
  \ket{\Phi_i} = \cre{c}{\alpha_{i,1}}\cdots\cre{c}{\alpha_{i,A}}\ket{},
\end{equation}
and is an eigenstate of a two-component $A$-body  ($A=N+Z$) fermionic \HO{}
Hamiltonian with the corresponding eigenenergy $E_i =
\left( N_i +\frac{3}{2} \right) \hw$ where $\Omega$ is the
oscillator frequency and $N_i$ is the total \HO{} energy quantum
number
\begin{equation}
  N_i \equiv \sum_{n_j,l_j \in \Phi_i} \left(2n_j+l_j \right),
\end{equation}
where $n_j$($l_j$) is the principle quantum number
(orbital angular momentum) of particle \(j\) in the basis state
$\Phi_i$.
The dimension \(D\) of the \NCSM{} basis is set by a total \HO{}-energy
truncation
\begin{equation}
  N_i - N_\mathrm{ref} \leq N_{\max},
\end{equation}
where $N_\mathrm{ref}$ is the total \HO{} energy quantum number of a
reference state composed of the $(N,Z)$ lowest
single-particle \HO{} states. For example, $N_\mathrm{ref} = 0$ for
\nuc{4}{He} and $N_\mathrm{ref} = 12$ for \nuc{16}{O}.

In general, there is no guarantee that the separation of intrinsic and
\CoM{} excitations due to the translational invariance of the
Hamiltonian is preserved when the Hilbert space is arbitrarily
truncated. However, an important feature of the total-energy
truncation of the \NCSM{} basis is that it does in fact guarantee this
separation due to the energy-conserving property of the \HO{}
transformation brackets~\cite{barrett2013}.
This property implies that there exists a unitary mapping of a \SD{}
basis of \HO{} single-particle states---truncated with respect to the
\HO{} excitation energy $\Nmax\hw$---onto a Jacobi-coordinate basis.
Therefore, the \NCSM{} eigenstates of~\eqref{eq:hamiltonian} can
formally be written as product states
\begin{equation}
  \ket{\Psi_n}_{\NCSM} =
  \ket{\Psi_{i}}_{\rm{int}} \otimes \ket{\Psi_{j}}_{\CoM},
  \label{eq:separable_state}
\end{equation}
with the state number $n = n(i,j)$. The lowest energy state ($n=0$)
will be the product of the ground state of the \CoM{} motion ($j=0$)
and that of the intrinsic Hamiltonian ($i=0$).
\subsection{Single-reference normal ordering}

In this section we outline the major steps of single-reference normal
ordering and describe the \NOTB{} approximation. Starting from the
general expression of the vacuum normal-ordered \threeNF{}s in
equation \eqref{eq:3nf_vac_no} and a reference state, that is a single
\SD{}
%
% Reference state
\begin{equation}
  \ket{\psi_{\rm{ref}}} =
  \cre{c}{\alpha_{1}}\cdots\cre{c}{\alpha_{A}}\ket{},
  \label{eq:reference_state}
\end{equation}
constructed from \(\mathcal{R} = \mathcal{R}(A,Z)=\{\alpha_{i}\}_{i=1}^A\)---the lowest \HO{}
states for the $A$-body system composed of $Z$ protons and $N=A-Z$ neutrons.

\(\hat{V}_{\rm{\threeNF}}\) can then be normal-ordered relative to
\(\ket{\psi_{\rm{ref}}}\), which results in an expansion of zero-,
one-, two- and three-body operators. The contribution to the
ground-state energy of the residual three-nucleon operator is assumed
to be small---since it acts solely outside the reference state---and
is discarded. This is known as the \NOTB{} approximation as it results
in an effective Hamiltonian with at most two-body operators.

% Contraction rules
The normal-ordering relative to \(\ket{\psi_{\rm{ref}}}\) is easiest performed with
Wick's theorem \cite{dickhoff2008}.
A product of second-quantization operators, normal-ordered relative to
\(\ket{\psi_{\rm{ref}}}\), is here written as
\(\NO{\hat{a}\hat{b}\hat{c}\cdots}\).
Such a normal-ordered operator fulfills
\( \NO{\hat{a}\hat{b}\hat{c}\cdots} \ket{\psi_{\rm{ref}}} = 0\). Combined with the formal
definition of a contraction,
\(\contraction{}{\hat{a}}{}{\hat{b}}\hat{a}\hat{b} = \hat{a}\hat{b} -
\NO{\hat{a}\hat{b}}\), it is possible to derive the contraction rules
\begin{align}
  \contraction{}{\cre{c}{\alpha}}{}{\anh{c}{\beta}}\cre{c}{\alpha}\anh{c}{\beta} &=
  \left\{
  \begin{array}{lc}
    \delta_{\alpha,\beta} & \text{if}\;\alpha \in \mathcal{R}(A,Z) \land \beta \in \mathcal{R}(A,Z)\\
    0 & \text{otherwise}
  \end{array} \right.\\
  \contraction{}{\anh{c}{\alpha}}{}{\cre{c}{\beta}}\anh{c}{\alpha}\cre{c}{\beta} &=
  \left\{
  \begin{array}{lc}
    \delta_{\alpha,\beta} & \text{if}\;\alpha \not\in \mathcal{R}(A,Z) \land \beta \not\in \mathcal{R}(A,Z)\\
    0 & \text{otherwise}
  \end{array} \right.\\
  \contraction{}{\cre{c}{\alpha}}{}{\cre{c}{\beta}}\cre{c}{\alpha}\cre{c}{\beta} &= 0\\
  \contraction{}{\anh{c}{\alpha}}{}{\anh{c}{\beta}}\anh{c}{\alpha}\anh{c}{\beta} &= 0.
\end{align}

The \threeNF{} in equation \eqref{eq:3nf_vac_no} can now be normal ordered
relative to \(\ket{\psi_{\rm{ref}}}\) by applying Wick's theorem,
%
% The normal orderd expression
%
\begin{equation}
  \label{eq:3nf_wick}
  \begin{aligned}
    \hat{V}_{\rm{\threeNF}} &=
    \overbrace{\sfrac{1}{6}\sum_{\alpha,\beta ,\gamma \in \mathcal{R}}
	    \sbra{\alpha\beta \gamma} V_{\rm{\threeNF}}\sket{\alpha\beta \gamma}}^{\equiv W_0}\\
    &\kern-1em +\overbrace{
	    \sfrac{1}{2}\sum_{\substack{\alpha\\ \alpha'}}\sum_{\beta
              ,\gamma \in \mathcal{R}}
	    \sbra{\alpha \beta  \gamma}{V}_{\rm{\threeNF}}\sket{\alpha' \beta  \gamma}
	    \NO{\cre{c}{\alpha}\anh{c}{\alpha'}}}^{\equiv \hat{W}_1}\\
    &\kern-1em +\overbrace{
	    \sfrac{1}{4}\sum_{\substack{\alpha\beta \\\alpha'\beta'}}
	    \sum_{\gamma \in \mathcal{R}}
	    \sbra{\alpha\beta \gamma}{V}_{\rm{\threeNF}}\sket{\alpha'\beta' \gamma}
	    \NO{\cre{c}{\alpha}\cre{c}{\beta}\anh{c}{\beta'}\anh{c}{\alpha'}}
    }^{\equiv \hat{W}_2}\\
    &\kern-1em +\overbrace{
	    \sfrac{1}{36}
	    \sum_{\substack{\alpha\beta\gamma \\ \alpha'\beta'\gamma'}}
	    \sbra{\alpha\beta\gamma}{V}_{\rm{\threeNF}}\sket{\alpha'\beta'\gamma'}
	    \NO{
		    \cre{c}{\alpha}\cre{c}{\beta}\cre{c}{\gamma}
		    \anh{c}{\gamma'}\anh{c}{\beta'}\anh{c}{\alpha'}
	    }
	    }^{\equiv \hat{W}_3},
  \end{aligned}
\end{equation}
where we note that $W_0$ is a constant while $\hat{W}_i \ket{\psi_{\rm{ref}}} = 0$ for $i \in
\{1,2,3\}$ due to the normal-ordered second-quantization operators.
%
% The approximation
The \NOTB{} approximation of $\hat{V}_{\rm{\threeNF}}$ is then defined as
\begin{equation} \label{eq:3nf_no2b_ref_no}
    \hat{V}^{\NOTB}_{\rm{\threeNF}}  \equiv W_0 + \hat{W}_1 + \hat{W}_2.
\end{equation}
In the \NCSM{}, however, the Hamiltonian is not expressed relative to a
reference state and we need to apply Wick's theorem backwards
to transform
\( \hat{V}^{\NOTB}_{\rm{\threeNF}} \) into vacuum
normal-ordered form. With this aim, we use the following relations
\begin{align}
 &\qquad\begin{aligned}
    \mathllap{\NO{\cre{c}{\alpha}\anh{c}{\alpha'}}} &=
	  \cre{c}{\alpha}\anh{c}{\alpha'} -
	  \contraction{}{\cre{c}{\alpha}}{}{\anh{c}{\alpha'}}
	  \cre{c}{\alpha}\anh{c}{\alpha'}
  \end{aligned}\\
  &\qquad\begin{aligned}
    \mathllap{
	    \NO{\cre{c}{\alpha}\cre{c}{\beta}\anh{c}{\beta'}\anh{c}{\alpha'}}}
    &=\cre{c}{\alpha}\cre{c}{\beta}\anh{c}{\beta'}\anh{c}{\alpha'}\\
    &\phantom{=}-\contraction{}{
      \cre{c}{\beta}
    }{}{
      \anh{c}{\beta'}
    }
    \cre{c}{\beta}\anh{c}{\beta'}
    \NO{
      \cre{c}{\alpha}\anh{c}{\alpha'}
    }
    -\contraction{}{
        \cre{c}{\alpha}
      }{}{
        \anh{c}{\alpha'}
      }
      \cre{c}{\alpha}\anh{c}{\alpha'}
    \NO{
      \cre{c}{\beta}\anh{c}{\beta'}
    }\\
    &\phantom{=}+
    \contraction{}{
      \cre{c}{\beta}
    }{}{
      \anh{c}{\alpha'}
    }
    \cre{c}{\beta}\anh{c}{\alpha'}
    \NO{
      \cre{c}{\alpha}\anh{c}{\beta'}
    }
    +
    \contraction{}{
      \cre{c}{\alpha}
    }{}{
      \anh{c}{\beta'}
    }
    \cre{c}{\alpha}\anh{c}{\beta'}
    \NO{
      \cre{c}{\beta}\anh{c}{\alpha'}
    }\\
    &\phantom{=}-
    \contraction{}{
      \cre{c}{\alpha}
    }{}{
      \anh{c}{\alpha'}
    }
    \cre{c}{\alpha}\anh{c}{\alpha'}
    \contraction{}{
      \cre{c}{\beta}
    }{}{
      \anh{c}{\beta'}
    }
    \cre{c}{\beta}\anh{c}{\beta'}
    +
    \contraction{}{
      \cre{c}{\alpha}
    }{}{
      \anh{c}{\beta'}
    }
    \cre{c}{\alpha}\anh{c}{\beta'}
    \contraction{}{
      \cre{c}{\beta}
    }{}{
      \anh{c}{\alpha'}
    }
    \cre{c}{\beta}\anh{c}{\alpha'}
  \end{aligned}
\end{align}
and arrive at
%
% The normal orderd expression
%
\begin{equation} \label{eq:3nf_no2b_vac_no}
  \begin{aligned}
    \hat{V}_{\rm{\threeNF}}^{\NOTB} &=
    \sfrac{1}{6}\sum_{\alpha,\beta ,\gamma \in \mathcal{R}}
	    \sbra{\alpha\beta \gamma} V_{\rm{\threeNF}}\sket{\alpha\beta \gamma}\\
    &\kern-1em -
	    \sfrac{1}{2}\sum_{\substack{\alpha\\ \alpha'}}\sum_{\beta
              ,\gamma \in \mathcal{R}}
	    \sbra{\alpha \beta  \gamma}{V}_{\rm{\threeNF}}\sket{\alpha' \beta  \gamma}
	    \cre{c}{\alpha}\anh{c}{\alpha'} \\
    &\kern-1em +
	    \sfrac{1}{4}\sum_{\substack{\alpha\beta \\\alpha'\beta'}}
	    \sum_{\gamma \in \mathcal{R}}
	    \sbra{\alpha\beta \gamma}{V}_{\rm{\threeNF}}\sket{\alpha'\beta' \gamma}
	    \cre{c}{\alpha}\cre{c}{\beta}\anh{c}{\beta'}\anh{c}{\alpha'}
            .
  \end{aligned}
\end{equation}
In the end, the \NOTB{}-approximated Hamiltonian that we use in the
\NCSM{} is
\begin{equation}
  \hat{H}^{\NOTB}=
	\hat{T}_{\rm{int}}+\hat{V}_{\rm{\twoNF}}+\hat{V}^{\NOTB}_{\rm{\threeNF}}.
  \label{eq:H_NO2B}
\end{equation}
This is clearly different compared to the full Hamiltonian, $\hat{H}$,
given by Eq.~\eqref{eq:hamiltonian} where the complete \threeNF{} is
retained.
%
%=====================================
\section{The center-of-mass problem%
  \label{sec:the-cm-problem}}
% ===
The translational symmetry of the \NOTB{}-approximated
Hamiltonian~\eqref{eq:H_NO2B} is explicitly broken since we neglect
the residual \threeNF{} and this renders the \CoM{} dependence of the
reference state manifest. Indeed, retaining the residual \threeNF{}
restores translational symmetry since the normal ordering in
Eq.~\ref{eq:3nf_wick} is an exact relation. With the \NOTB{}
approximation it is therefore no longer guaranteed that the ground
state $\ket{\Psi^{\NOTB}_{\rm{gs}}}_{\NCSM}$ of
\(\hat{H}^{\rm{\NOTB}}\) is factorized into a product of \CoM{} and
intrinsic states as in equation \eqref{eq:separable_state}. Instead we
must expect a linear superposition of product states,
\begin{equation}
\ket{\Psi^{\rm{\NOTB}}_{\rm{gs}}}_{\NCSM} =
\sum_{i,j}c_{i,j}\ket{\Psi_{i}}_{\rm{int}} \otimes \ket{\Psi_{j}}_{\CoM}.
\end{equation}
In this more general situation the intrinsic and \CoM{}
states are no longer pure quantum states and must be expressed with density
matrices \(\hat{\rho}_{\rm{int}}\) and \(\hat{\rho}_{\rm{\CoM}}\).

This mixing of \CoM{} and intrinsic degrees of freedom can potentially
have a huge effect on various observables and is here labeled as the
\textit{center-of-mass problem}. It is therefore crucial to quantify
the \CoM{} mixing. In the following we will introduce two metrics that
have exactly this purpose.
%
%-------------
\subsection{Introducing center-of-mass metrics%
  \label{sec:CMmetrics}
  }
%-------------
%
  The mixing of \CoM{} and intrinsic states is a known problem in
  many-body physics. It might occur also when using fully
  translational-invariant Hamiltonians as a consequence of
  approximations used in the many-body solver. In particular, Galilean
  invariance is broken explicitly when employing lattice
  methods~\cite{lee2009} and \CoM{} mixing can occur in basis-expansion
  methods when imposing a basis truncation at the
  single-particle level~\cite{gloeckner1974,hagen2009a, roth2009b,
    hergert2016, parzuchowski2017, tichai2018, hoppe2020}. A very
  common approach to diagnose the problem in basis expansion methods
  is to evaluate the (energy-shifted) \HO{} \CoM{} Hamiltonian
\begin{equation}
  \hat{H}_{\rm{\CoM}}(\omega)=
	\frac{\hat{P}_{\rm{\CoM}}^2}{2 m A}+\frac{1}{2}mA\omega^2\hat{R}_{\rm{\CoM}}^2-
	\frac{3}{2}\hbar\omega,
   \label{eq:HCM}
 \end{equation}
 or the corresponding \CoM{} number operator
 \begin{equation}
  \hat{N}_{\rm{\CoM}}(\omega) = \frac{1}{\hbar\omega}
  \hat{H}_{\rm{\CoM}}(\omega)
  \label{eq:NCM}
\end{equation}
with expectation value \(\NCM{}(\omega)\).
Small expectation values of these operators,
$\hat{H}_{\rm{\CoM}}(\Omega)$ and $\NCM{}(\Omega)$, evaluated at the
basis frequency $\Omega$, are then used as evidence for satisfactory
\CoM{} factorization. Large expectation values, on the other hand,
indicate problematic mixing.

However, it might be too assertive to claim proper \CoM{} separation
based on this single observable. We argue here that additional metrics
are needed. In addition, there are claims~\cite{hagen2009a} that the factorization does
occur but that the \CoM{} state is not necessarily a ground state of
the Hamiltonian $\hat{H}_{\rm{\CoM}}(\Omega)$~\eqref{eq:HCM}
constructed using the basis frequency \hw.
%
%=========
\subsubsection{The \xiCM{} metric}
% ===
Consider an eigenstate of a translationally invariant Hamiltonian that
factorizes into a product of an intrinsic state $\ket{\Psi}_{\rm int}$
and a \CoM{} state $\ket{\Psi_{\rm{gs}}^{\wxi}}_{\CoM}$, where the
latter corresponds to the ground state of
$\hat{H}_{\rm{\CoM}}(\wxi)$---the Hamiltonian~\eqref{eq:HCM}
constructed with an oscillator frequency $\wxi$ which does not
necessarily correspond to the basis frequency $\Omega$. In this
situation we would obtain the expectation values
\begin{align}
    \expect{ R^2_{\rm{\CoM}}} &= \frac{3}{2} b^2 \\
\intertext{and}
  \expect{ P^2_{\rm{\CoM}}} &= \frac{3}{2}\frac{\hbar^2}{b^2},
\end{align}
with the oscillator length $b = b(\wxi) = \sqrt{\hbar / A m \wxi}$,
and the expectation values are with respect to the full ground state.
This fact was utilized by \textcite{parzuchowski2017} in their study
of transition operators within the in-medium SRG framework. They
introduced the quantity
\begin{equation}
  \xiCM \equiv
	\frac{\sqrt{\langle \hat{R}^2_{\rm{\CoM}}\rangle\langle
	\hat{P}^2_{\rm{\CoM}}\rangle}}{\hbar}-\frac{3}{2},
\end{equation}
which will evaluate to $\xiCM = 0$ if \(\ket{\Psi_{\rm{gs}}^{\wxi}}_{\CoM}\)
is a \HO{} ground state, regardless of the frequency $\wxi$, while
\(\xi_{\rm{\CoM}} > 0\) if it is not.
Note, however, that a \HO{} eigenstate
with one frequency, \(\omega\), cannot be exactly represented in a
truncated \HO{} basis with a different basis frequency \(\Omega \neq
\omega\).  This is
illustrated in Fig.~\ref{fig:ho_representation_xi} and further
discussed in Appendix~\ref{sec:representations}.

In the case when \(\xi_{\rm{\CoM}}\approx 0\) it is possible to identify the
corresponding frequency of the underlying \HO{} Hamiltonian
$\hat{H}_{\CoM}(\wxi)$ by
\begin{equation}\label{eq:omega_xi}
  \hbar\wxi = \frac{4}{3}\expect{\hat{T}_{\rm{\CoM}}},
\end{equation}
where \(\hat{T}_{\rm{\CoM}}\) is the \CoM{} kinetic energy, and the expectation
value is with respect to the ground state. When the \NCSM{} Hamiltonian is
translationally invariant the frequency \(\wxi\) will equal the basis frequency
in the \NCSM{} method. However, \(\wxi \neq \Omega\) indicates a broken
symmetry. Note that we might still have a product
state~\eqref{eq:separable_state} in this situation---such that \CoM{} mixing is not
problematic---and that we can measure this with \(\xiCM{}\).
\begin{figure}
  \includegraphics[width=\columnwidth]{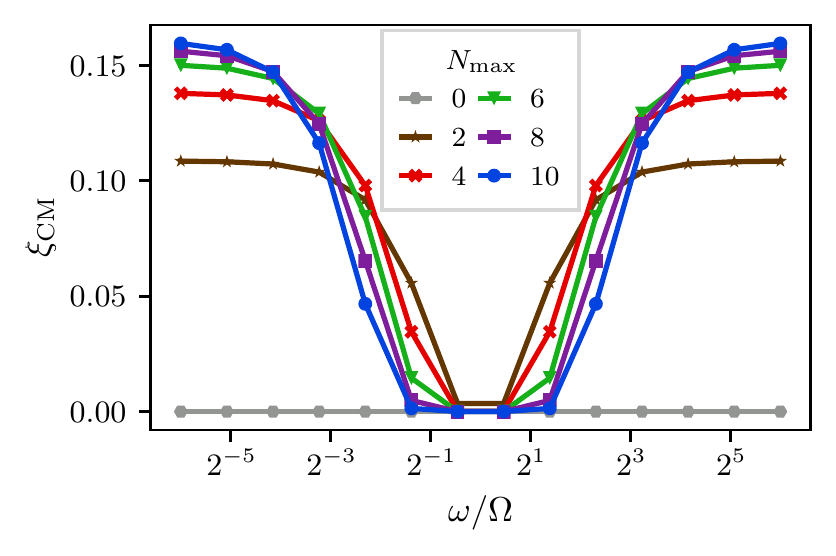}
  \caption{
	  \label{fig:ho_representation_xi} The \xiCM{} metric for
	  ground states of \(H_{\rm{\CoM}}(\omega)\) computed for
          different $\omega/\Omega$ ratios (where $\Omega$ is the basis
          frequency) and increasing \NCSM{} basis truncations. Note
          that $\xiCM{} \to 0$ for all $\omega/\Omega$ ratios as
          $\Nmax \to \infty$.
	}
\end{figure}
%
%=========
\subsubsection{The \NCM{} metric}
%===
%
In applications of the \CC{} method the computed ground state
\(\ket{\Psi_{\rm{gs}}}_{\rm{CC}}\) is assumed to be separable
such that the \CoM{} state is the ground state of a \HO{} Hamiltonian with a frequency
\(\wN\) that not necessarily equals the underlying \HO{} basis
frequency \(\Omega\).
The frequency \(\wN\) is obtained by evaluating~\cite{hagen2009a}
\begin{multline}
\hbar\omega^\pm =
\hbar\Omega+\frac{2}{3}\expect{\hat{H}_{\rm{CM}}(\Omega)} \\
\pm\sqrt{\frac{4}{9}\expect{\hat{H}_{\rm{CM}}(\Omega)}^2
  + \frac{4}{3}\hbar\Omega\expect{\hat{H}_{\rm{CM}}(\Omega)}},
\end{multline}
and identifying
\begin{equation}
  \wN = \argmin_{\omega \in \omega^\pm} \expect{\hat{H}_{\rm{CM}}(\omega)}.
\end{equation}
We can then define the operator $\hat{N}_{\CoM}(\wN)$ analogous to
Eq.~\eqref{eq:NCM} and evaluate its expectation value
\(\NCM{}(\wN)\) which will be small if the wave function
factorizes.
%
% ===============
\subsection{The relation between different metrics}\label{sec:relation_between_the_metrics}
% ===
There is an interesting connection between \(\xiCM{}\) and
\(\NCM{}(\wxi{})\), i.e. the two different metrics expressed in terms
of the same \CoM{} oscillator frequency $\wxi{}$. We pick \(\wxi\) since the
two frequencies \(\wxi\) and \(\wN\) are equal in the limit
\(\Nmax \to \infty\) if the state is separable as in equation
\eqref{eq:separable_state}.
Consider a factorized state with
\begin{equation}
\ket{\Psi}_{\rm{\CoM}} =
\ket{\phi^{\wxi}_{\mathcal{N}\mathcal{L}}},
\end{equation}
i.e., it is a \HO{} state with frequency \(\wxi\), radial quantum number \(\mathcal{N}\) and
orbital angular momentum \(\mathcal{L}\). Then we have
\begin{equation}
  \NCM{}(\wxi{}) = 2\mathcal{N}+\mathcal{L},
\end{equation}
\begin{align}
\langle R^2_{\rm{\CoM}}\rangle &=
	b(\wxi)^2\left( 2\mathcal{N}+\mathcal{L}+\frac{3}{2} \right), \\
\intertext{and}
\langle P^2_{\rm{\CoM}}\rangle &=
	\frac{\hbar^2}{b(\wxi)^2}\left(
                                 2\mathcal{N}+\mathcal{L}+\frac{3}{2} \right).
\end{align}
Using the definitions of the two metrics we find that they become
equal in this scenario
\begin{equation}
  \xiCM{} = \frac{\sqrt{\langle \hat{R}^2_{\rm{\CoM}}\rangle\langle
	\hat{P}^2_{\rm{\CoM}}\rangle}}{\hbar}-\frac{3}{2} =
	2\mathcal{N}+\mathcal{L} = \NCM{}(\wxi{}).
\end{equation}

Now consider the possibility that $\ket{\Psi}_{\rm{\CoM}}$ is a
linear superposition
\begin{equation}
  \ket{\Psi}_{\rm{\CoM}} = \sum_{i} c_i \ket{\phi_{\mathcal{N}_i\mathcal{L}_i}},
\end{equation}
of \HO{} states $\ket{\phi_{\mathcal{N}_i\mathcal{L}_i}} = a^\dagger_{\mathcal{N}_i\mathcal{L}_i}\ket{}$.
To simplify further calculations we introduce
\begin{align}
  A=&\sum_{i}|c_i|^2\left(2\mathcal{N}_i+\mathcal{L}_i+\frac{3}{2}\right)\\
    \intertext{and}
  B=&\sum_{i,j}c^*_ic_j(\sqrt{\mathcal{N}_i(\mathcal{N}_i+\mathcal{L}_i+1/2)}\delta_{\mathcal{N}_i,\mathcal{N}_j+1}\nonumber\\
       &+\sqrt{\mathcal{N}_j(\mathcal{N}_j+\mathcal{L}_i+1/2)}\delta_{\mathcal{N}_i+1,\mathcal{N}_j})\delta_{\mathcal{L}_i,\mathcal{L}_j}
\end{align}
where \(A, B\) are real and $A \ge 3/2$.
Then we find
\begin{equation}\label{eq:NCMbar}
  \NCM{}(\wxi{})  = A-\frac{3}{2},
\end{equation}
\begin{align}
    \langle R^2_{\rm{\CoM}}\rangle &= b^2(A-B) \\
    \intertext{and}
\langle P^2_{\rm{\CoM}}\rangle &= \frac{\hbar^2}{b^2}(A+B),
\end{align}
which gives
\begin{equation}\label{eq:XICM}
\xiCM{} = \sqrt{A^2-B^2}-\frac{3}{2}.
\end{equation}
It is clear from equations \eqref{eq:NCMbar} and \eqref{eq:XICM} that
\begin{equation}
  \NCM{}(\wxi{})-\xiCM{}=\frac{A}{2} \left(
    \varepsilon +\mathcal{O}(\varepsilon^2) \right),
  \label{eq:NCM_xiCM_relation}
\end{equation}
where we have assumed that $\varepsilon \equiv B^2/A^2 \ll
1$. Therefore, the difference between these two metrics can be used as
a measure of how much of the \CoM{} state is in higher excitations.
The off-diagonal sum \(B\)
can only be non-zero if there exists \(i,j\) such that \(c_i,c_j\neq
0\) with \(|\mathcal{N}_i-\mathcal{N}_j|=1\) and \(\mathcal{L}_i =
\mathcal{L}_j\).

It turns out that a similar relation can be derived if we have \CoM{}
mixing such that the \CoM{} state is not a pure quantum state. Then we
find that
\(\xiCM{} \neq \NCM{}\) if there does not exist any \HO{} basis in which the
\CoM{} density matrix
\begin{equation}
  (\rho_{\rm{CM}})_{i,j} = \bra{\Psi _\mathrm{NCSM}}
  {a^\dagger_{\mathcal{N}_i\mathcal{L}_i }
    a_{\mathcal{N}_j\mathcal{L}_j }} \ket{\Psi _\mathrm{NCSM}}, 
\end{equation}
is diagonal.
In this situation the coefficients $A$ and $B$ in Eqs.~\eqref{eq:NCMbar} and 
\eqref{eq:XICM} are given by
\begin{align}
  A=&\sum_{i}(\rho_{\rm{\CoM}})_{i,i}\left(2\mathcal{N}_i+\mathcal{L}_i+\frac{3}{2}\right)\\
    \intertext{and}
  B=&\sum_{i,j}(\rho_{\rm{\CoM}})_{i,j}(\sqrt{\mathcal{N}_i(\mathcal{N}_i
      +\mathcal{L}_i+1/2)}\delta_{\mathcal{N}_i,\mathcal{N}_j+1}\nonumber\\
       &+\sqrt{\mathcal{N}_j(\mathcal{N}_j+\mathcal{L}_i+1/2)}\delta_{\mathcal{N}_i+1,\mathcal{N}_j})
         \delta_{\mathcal{L}_i,\mathcal{L}_j}.
\end{align}
In conclusion, when finding that \(\xiCM{}, \NCM{}>0\) we cannot know
if the \CoM{} state is a pure quantum state or a mixed one. Only the
situation \(\xiCM{} = \NCM{} = 0\) assures a proper separation of the
intrinsic and \CoM{} parts of the eigenstate as in Eq.~\eqref{eq:separable_state}.
%
%-------------
\subsection{Benchmark of \CoM[fullname]{} metrics}
%-------------

To benchmark the \CoM{} analysis metrics we consider an interacting
many-body system in an external \HO{} trap with Hamiltonian
\begin{equation}
\hat{H}^{\rm{trap}} =
\hat{T}_{\rm{int}}+\hat{V}_{\rm{\twoNF}}+\hat{H}_{\rm{\CoM}}(\wtrap),
\label{eq:Htrap}
\end{equation}
where we use the \twoNF{} part of \NNLOsat{} as a realistic
interaction $\hat{V}_{\rm{\twoNF}}$.
Then we compute the \NCSM{} ground state of \nuc{4}{He} for different
basis frequencies \(\hw \in \{8, 12,\ldots,32,36\}\;\unit{MeV}\) while
keeping the trapping potential frequency fixed at
$\hbar\wtrap = 20$~MeV. The metrics described in
Sec.~\ref{sec:CMmetrics} are then evaluated for the ground state.

While this Hamiltonian depends on the \CoM{} coordinate---such that
translational invariance is explicitly broken---it is still
block-diagonal in a \CoM{}-part and an intrinsic part. Therefore, it is possible
to precisely control the \CoM{} part of the ground state and
this property makes it a suitable benchmark of the \CoM{}
metrics. However, it is not equivalent to the non block-diagonal
\CoM{} coupling of the \NOTB{}-approximated Hamiltonian.

\begin{figure*}[htb]
  \begin{center}
  \includegraphics[width=\textwidth]{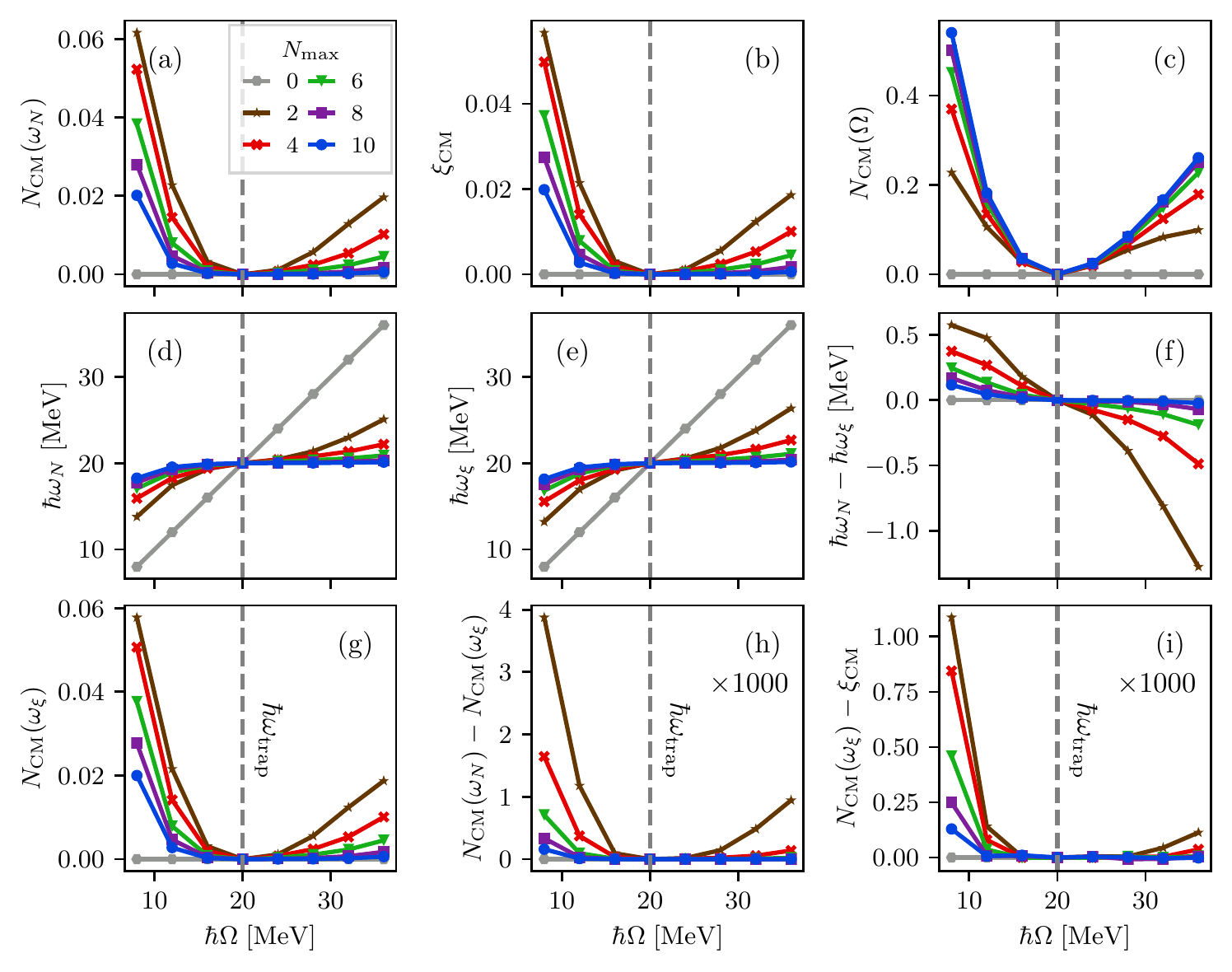}
  \caption{
	  \label{fig:breaking_translational_symmetry_4He}
	  \CoM{}-excitation metrics and \HO{} frequencies of
          \nuc{4}{He} computed with an external trap
 	  \(\hat{H}^{\rm{trap}}\)~\eqref{eq:Htrap} with \(\hbar\wtrap= 20\;\unit{MeV}\)
          using different basis frequencies $\hw{}$. Note in particular
          that the two metrics \NCM{} and \xiCM{}---shown in
          panels (a,b), respectively---are very similar and
          that both of them correctly identify the trap frequency for
          a wide range of basis frequencies---see panels (d,e). In contrast, the standard metric
          $\NCM(\Omega)$, shown in panel (c), does not reveal the
          actual decoupling except for $\Omega = \wtrap$. The
          differences shown in panels (h,i) are multiplied by a factor
          1000.
  }
  \end{center}
\end{figure*}

We compute the expectation values $\expect{ R^2_{\rm{\CoM}}}$,
$\expect{ P^2_{\rm{\CoM}}}$ and $\expect{\hat{H}_{\rm{CM}}(\omega)}$
for each \NCSM{} model space $(\Nmax,\hw)$. This allows us to extract the optimal
decoupling frequencies $\wxi$ and $\wN$ and to test the decoupling by
evaluating the metrics $\xiCM$ and $\NCM(\wN)$. In addition, the standard
\CoM{}-decoupling metric $\NCM(\Omega)$ can be evaluated---although it
is expected to fail when $\wtrap \neq \Omega$. All of these quantities
are plotted in Fig.~\ref{fig:breaking_translational_symmetry_4He}.

We numerically confirm that all three metrics, shown in the top row of
Fig.~\ref{fig:breaking_translational_symmetry_4He}, become equal to
zero when the basis frequency $\Omega$ is equal to the trap
frequency. However, while $\NCM(\Omega)$ in panel (c) fails to reveal
the decoupling for other basis frequencies, the two metrics
$\NCM(\wN)$ and $\xiCM$, shown in panels (a) and (b), respectively,
do indicate decoupling by exhibiting small values.

The fact that both $\NCM(\wN)$ and $\xiCM$ are larger for small basis
frequencies indicates that a superposition of excited \HO{} states is needed to
describe the \CoM{} ground state in this truncated space (see
Appendix~\ref{sec:representations}).
However, for \(\hw > \hbar \wtrap\) the metrics are very small already
at modest \Nmax{} indicating a good \CoM{} state representation with these basis
frequencies.

The corresponding optimal frequencies, \(\hbar\wN{}\) and
\(\hbar\wxi{}\) shown in panels (d) and (e), do approach the trap
frequency as \Nmax{} increases. For $\Nmax = 0$, where there is a
single \SD{} basis state, this analysis will always return the basis
frequency as the optimal one, as shown by the diagonal, straight line.

Finally, the differences between the optimal frequencies found via the
\xiCM{} and \NCM{} methods are shown in panel (f), while the
difference between the two metrics are displayed in panels (h) and
(i). As a general conclusion we find that the two analysis methods
provide basically identical results, but that the \xiCM{} metric is
easier to implement and compute. Furthermore, the problem of 
representing a \HO{} state of another frequency than that of the truncated 
\HO{} basis hampers the analysis at small basis frequencies (see
Appendix~\ref{sec:representations}). 
%
%=====================================
\section{\NOTB{} results %
\label{sec:results}}
% -------------
%
In this section we present a numerical study of the \SRNOTB{}
approximation in the \HO{} \SD{} basis applied to the doubly-magic
systems \nuc{4}{He} and \nuc{16}{O}. All results shown here are
obtained with the realistic nuclear interaction model
\NNLOsat~\cite{ekstrom2015a}. Throughout this study we will compare
results obtained with full and with \NOTB{}-approximated \threeNF{}s.

We employ the Jacobi-coordinate version of the
\NCSM{}~\cite{kamuntavicius2000,ekstrom2013} to compute the ground state of
\nuc{4}{He} with full \threeNF{}. The normal ordering is performed in
the $M$-scheme \SD{} basis and we employ the NCSM code
pAntoine~\cite{Navratil:2003ib,forssen2017} to perform the diagonalization.
Unfortunately, the huge number of \threeNF{} matrix elements in the
$M$-scheme \SD{} basis limits our studies to model spaces
$\Nmax \leq 10$. Specifically, with \(\Nmax = 10\) we have \(5.4 \cdot
10^9\) elements
while \(\Nmax = 12\) would require \(66.5 \cdot 10^9\).

For \nuc{16}{O} we are limited by the size and the number of
non-zero elements of the Hamiltonian
matrix. With \NOTB{}-approximated interactions we use pAntoine and
reach model spaces $\Nmax \leq 8$ with up to
$D = 6 \cdot 10^8$ basis states. With full \threeNF{}s we use the NCSD
code~\cite{NCSD} and are able to reach model spaces $\Nmax \leq 6$ corresponding
to $D = 1.6 \cdot 10^6$.

The direct comparison between results obtained with full and
\NOTB{}-truncated \threeNF{}s allows us to focus on the size of the
approximation error as a function of the mass number and model space
parameters. The origin of the approximation error will here be analyzed in
terms of possible \CoM{} mixing.
In this context it is important to point out that all calculations in
the $M$-scheme \SD{} basis are performed without a Lawson projection
term acting on the \CoM{} coordinates. Instead, we will employ the
metrics presented in Sec.~\ref{sec:CMmetrics} as diagnostic tools.
%
%-------------
\subsection{Ground state energy of \nuc{4}{He}}
\begin{figure}[ht!]
	\includegraphics[width=\columnwidth]{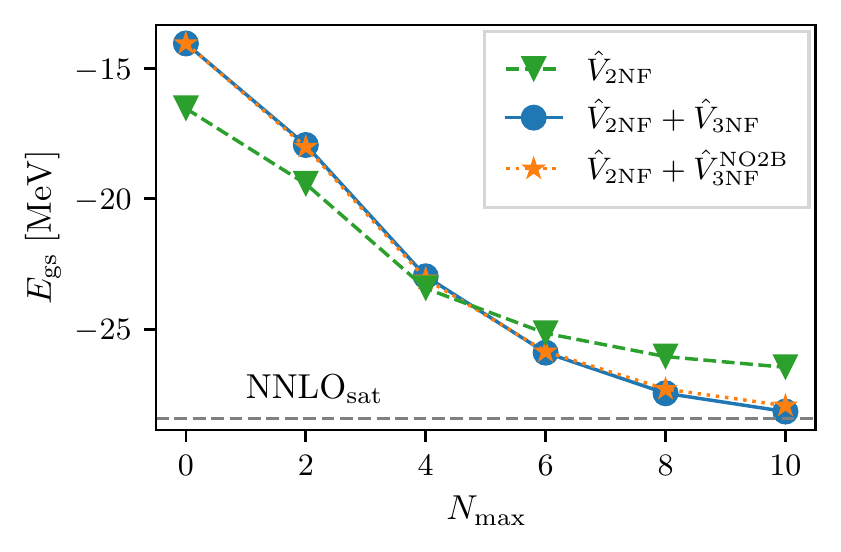}
	\caption{\label{fig:comp_4He_2nf_3nf_no2b_hw_20} The
          ground-state energy of \nuc{4}{He} computed with the
          \NNLOsat{} interaction for fixed basis frequency \(\hw =
          20\;\unit{MeV}\). Results with only the \twoNF{} interaction
          (green, dashed line) is compared with the full Hamiltonian
          including \threeNF{}s (blue, solid line) and with the
          \NOTB{}-approximated one (orange, dotted line),
          corresponding to Eqs.~\eqref{eq:hamiltonian} and
          \eqref{eq:H_NO2B}, respectively.  The dashed horizontal line
          indicates the converged \NNLOsat{} ground-state energy
          \(E_{\rm{gs}} = -28.43\;\unit{MeV}\)~\cite{ekstrom2015a}.  }
\end{figure}

%-------------
%
We first compute the ground-state energy of \nuc{4}{He} at the fixed
basis frequency \(\hw = 20\;\unit{MeV}\), which is close to the
position of the variational minimum for this system with the
\NNLOsat{} interaction. Results are shown in
Fig.~\ref{fig:comp_4He_2nf_3nf_no2b_hw_20} as a function of increasing
\NCSM{} truncation \(N_{\max}\) and compared to the converged result
\(E_{\rm{gs}}=-28.43\;\unit{MeV}\)~\cite{ekstrom2015a}.

At this basis frequency, we find that the \NOTB{} approximation
captures the \(N_{\max}\) behavior of the results obtained with the
full Hamiltonian to within \(1\%\). This means that the approximation
error in the total binding energy is smaller than \(250\;\unit{keV}\).
We can also observe the importance of the \threeNF{} since a full
removal of this part of the Hamiltonian (green dashed line in
Fig.~\ref{fig:comp_4He_2nf_3nf_no2b_hw_20}) leads to
approximately \(2\;\unit{MeV}\) underbinding.

\begin{figure}[ht!]
	\includegraphics[width=\columnwidth]{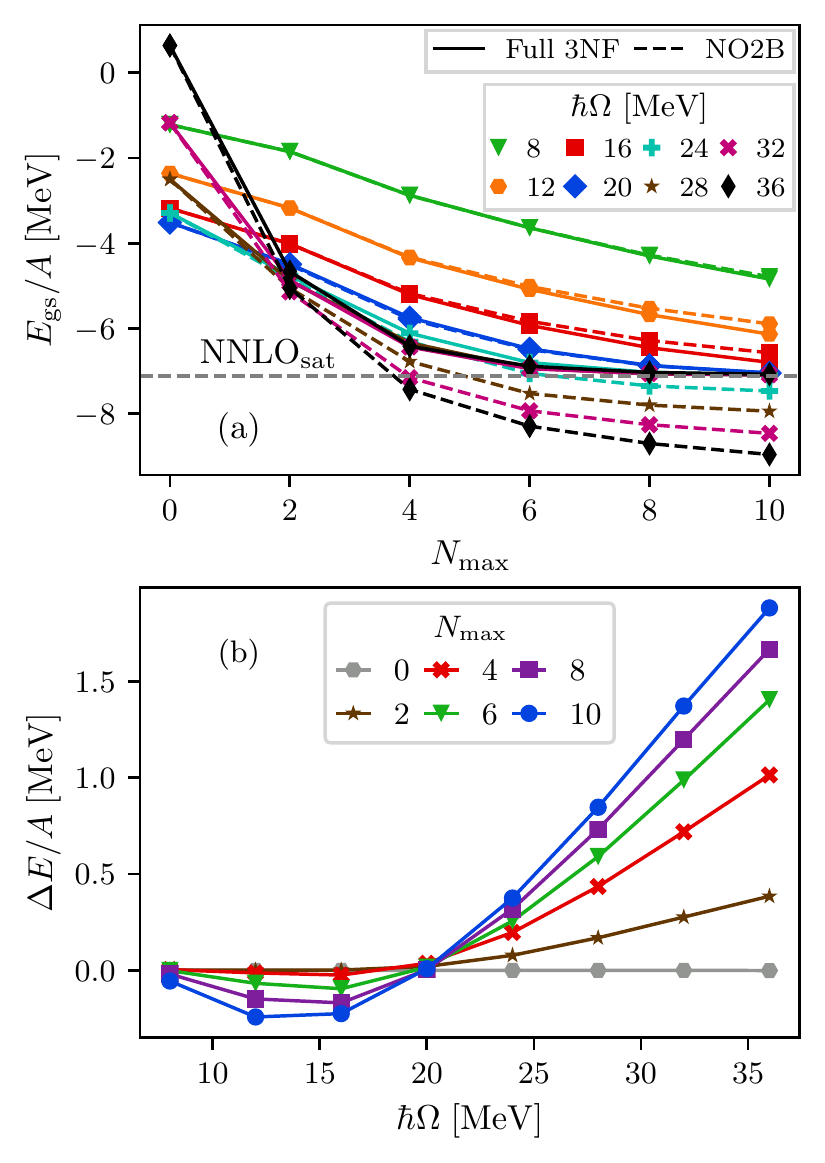}
	\caption{\label{fig:comp_4He_3nf_no2b_hw_4_to_36_step_4}
          (a) The ground-state energy per
          nucleon of \nuc{4}{He} computed with the \NNLOsat{}
          interaction for different \(\hw\). The solid lines show
          results with full inclusion of the \threeNF{}, while the
          dashed lines correspond to \NOTB{}-approximated
          \threeNF{}. Although the convergence rate is
          \hw{}-dependent, the full \threeNF{} results converge
          towards the exact result (horizontal dashed line) while the
          results with the \NOTB{} approximation do not. (b) The
          \NOTB{}-approximation error $\Delta E \equiv
          E_{\rm{gs}}^{\threeNF{}} - E_{\rm{gs}}^{\NOTB{}}$ per
          nucleon where $E_{\rm{gs}}^{\threeNF{}}$ and
          $E_{\rm{gs}}^{\NOTB{}}$ are the ground-state eigenenergies
          of Eq.~\eqref{eq:hamiltonian} and \eqref{eq:H_NO2B},
          respectively.  }
\end{figure}

However, the magnitude of the \NOTB{}-approximation error turns out to
be highly sensitive to the choice of basis frequency. This finding is
highlighted in Fig.~\ref{fig:comp_4He_3nf_no2b_hw_4_to_36_step_4}
where the binding energy per nucleon is computed for
\(\hw \in \{8, 12,\dots,32,36\}\;\unit{MeV}\). The solid
lines in the upper panel correspond to \(E_{\rm{gs}}/A\) computed with the full Hamiltonian, while
the dashed ones correspond to the \NOTB{}-approximated \threeNF{}. The
difference between these two results is shown in the lower panel as a
function of the basis frequency.

There seem to be an optimal frequency
\(\hw \approx 20\;\unit{MeV}\) for which the approximation error is
very small as we transition from under- to overbinding with the
\NOTB{} truncation.
For higher frequencies there is an increasing difference between the \NOTB{} and the
full-\threeNF{} results.
Note also that the \NOTB{} truncation at $\Nmax = 0$ is identical
to the full Hamiltonian as the single reference state is the only
basis state.

The main hypothesis of this paper is that the explicitly broken
translational symmetry of the \NOTB{} Hamiltonian can become the
origin of a strong \(\hw\)-dependence of the approximation
error. Consequently, the \NCSM{} eigenstates might
not necessarily separate into a product of \CoM{} and intrinsic
states.

\begin{figure}[ht!]
	\includegraphics[width=\columnwidth]{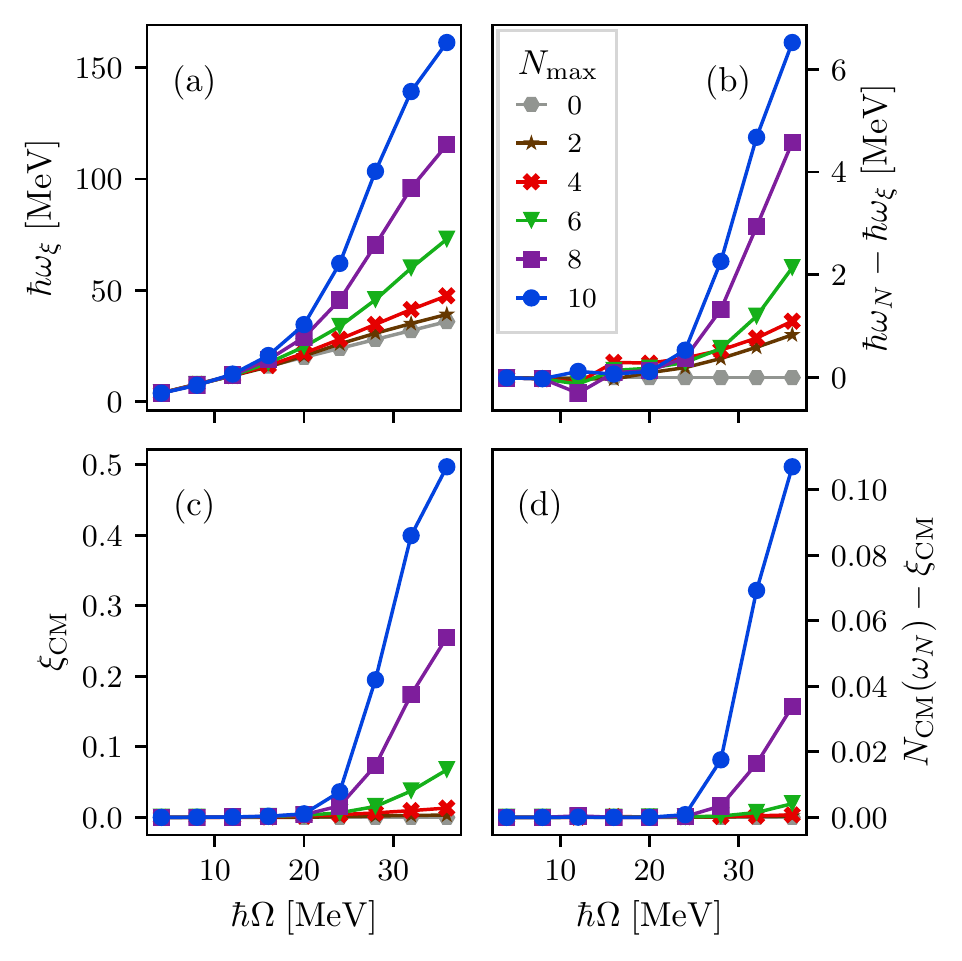}
	\caption{\label{fig:4he_center_of_mass_excitation_combined}
          \CoM{} analysis for \nuc{4}{He} eigenstates obtained with the \NOTB{} Hamiltonian. The frequency
          $\hbar\wxi$ from Eq.~\eqref{eq:omega_xi} is shown in panel
          (a) and the
          difference $\hbar\wN - \hbar\wxi$ is shown in panel (b). Note that the basis
          frequency $\hw$ is equal to both $\hbar\wN$ and $\hbar\wxi$
          at $\Nmax=0$, as shown
          by the gray line with circle markers.
          The \CoM{} metric \xiCM{} from Eq.~\eqref{eq:XICM} is shown
          in panel (c) and the
          difference $\NCM(\wN) - \xiCM$ in panel (d).
        }
\end{figure}

To test this hypothesis we evaluate the \CoM{} metrics, \xiCM{} and
\NCM{}---defined in Sec.~\ref{sec:CMmetrics}---and the corresponding
\CoM{} oscillator-state frequencies \wxi{} and \wN{}. These results
are shown in
Fig.~\ref{fig:4he_center_of_mass_excitation_combined}.
We observe that $\hbar\wxi \gg \hw$ for large basis
frequencies. However, the \xiCM{} metric clearly indicates that there
is no \CoM{} separation in this scenario so the value of \wxi{} does
not really have any significance.

In contrast, for small basis frequencies we have a clear factorization
of the eigenstate, as indicated by both metrics, and we also find that
the extracted frequencies are very similar and very close to the basis
frequency. There is a transition region around \(\hw \approx
20\;\unit{MeV}\) where the metrics indicate \CoM{} separation at a
frequency that is slightly larger than the basis one.

The finding that \CoM{} mixing is less of a concern for small basis
frequencies also indicates that the \NOTB{}-approximation error of
$\lesssim 500\;\unit{keV}/A$ in this region is due to the neglected,
residual \threeNF{}.
As shown in the lower panels of
Fig.~\ref{fig:4he_center_of_mass_excitation_combined} we find that
both \xiCM{} and \NCM{} do become very small for basis frequencies
below $\approx 20 \; \unit{MeV}$, indicating a separation between the
\CoM{} and intrinsic parts of \nuc{4}{He} ground state. However, as
the basis frequency increases beyond \(20 \;\unit{MeV}\) both measures
increase drastically, suggesting that there is no longer any separation.
The frequencies \(\hbar\omega_{N}\) and \(\hbar\omega_{\xi}\) start
to differ visibly from the basis frequency already at \(16 \MeV\)
which is below the observed optimal frequency. This indicates that the
\NOTB{} approximation does affect the \CoM{} state, albeit very weakly.
%
%-------------
\subsection{Ground state energy of \nuc{16}{O}}
%-------------
%
We will now study the \NOTB{} approximation when performing \NCSM{}
calculations of the \nuc{16}{O} nucleus. For this system we are
limited to $\Nmax \le 8$ for Hamiltonians including only \twoNF{}s,
and $\Nmax\le 6$ when using the Hamiltonian with full
\threeNF{}s. Such differences in computational limits are the main
reason for using the \NOTB{} approximation in the first place.
In this work, \NCSM{} computations with the full \threeNF{} for
\nuc{16}{O} are only performed at a few basis frequencies:
$\hw = 16, 20, 24, 36 \; \unit{MeV}$.

\begin{figure}[ht!]
	\includegraphics[width=\columnwidth]{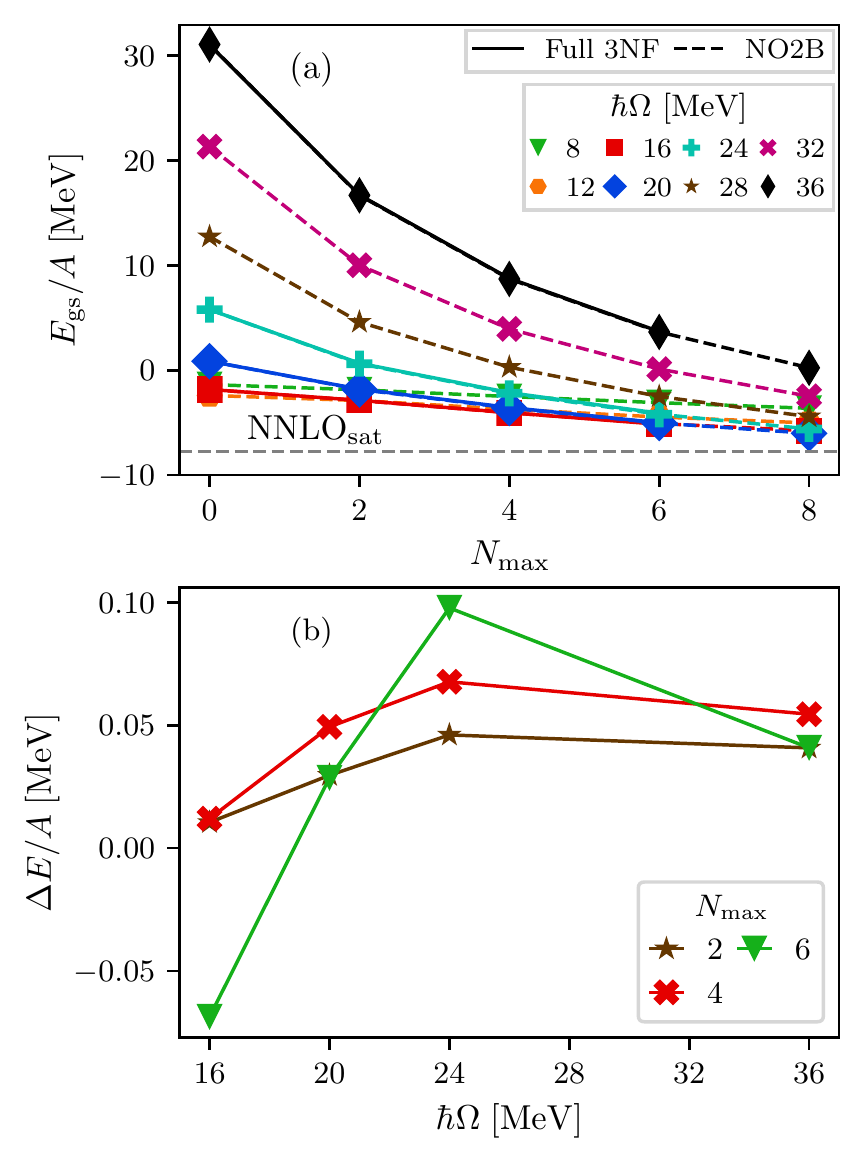}
	\caption{\label{fig:binding_energy_16O_3nf_vs_no2b_hw_4_to_36_step_4}
	(a) The ground-state energy per particle of \nuc{16}{O} computed with the \NNLOsat{} interaction
	for different \(\hw\). The solid lines show results with full
        inclusion of the \threeNF{} (only available for a subset of \HO{}
        frequencies and up to $\Nmax=6$), while the dashed lines correspond
        to \NOTB{}-approximated \threeNF{}. The horizontal dashed line
        is the converged
	N2LO\(_{\rm{sat}}\) result computed with the \CC{} method~\cite{ekstrom2015a}.
	(b)
        The \NOTB{}-approximation error $\Delta E \equiv
        E_{\rm{gs}}^{\threeNF{}}  - E_{\rm{gs}}^{\NOTB{}}$ per particle
        where $E_{\rm{gs}}^{\threeNF{}}$ and $E_{\rm{gs}}^{\NOTB{}}$ are the
        ground-state eigenenergies of Eq.~\eqref{eq:hamiltonian} and
        \eqref{eq:H_NO2B}, respectively.
      }
\end{figure}

In addition, it is well known that \CoM{} effects are suppressed in
heavier systems since the excitation of \CoM{} motion is energetically
costly. Accordingly, in
Fig.~\ref{fig:binding_energy_16O_3nf_vs_no2b_hw_4_to_36_step_4} we
find that the \NOTB{} approximation captures the \Nmax{} dependence of
the ground-state energy results rather well for a wide frequency
range. Note, however, that we are relatively far from convergence at
$\Nmax=8$ when using large basis frequencies. For comparison, we also show
the converged result from \CC{} calculations
\(E_\mathrm{gs} / A = -7.78 \; \unit{MeV}\)~\cite{ekstrom2015a}.

The lower panel of
Fig.~\ref{fig:binding_energy_16O_3nf_vs_no2b_hw_4_to_36_step_4} shows
that the \NOTB{}-approximation error is on the order of
$\lesssim 100\;\unit{keV}/A$, corresponding to $\sim 1.5 \; \unit{MeV}$ in the
total binding energy (just over 1\%).

\begin{figure}[htb]
	\includegraphics[width=\columnwidth]{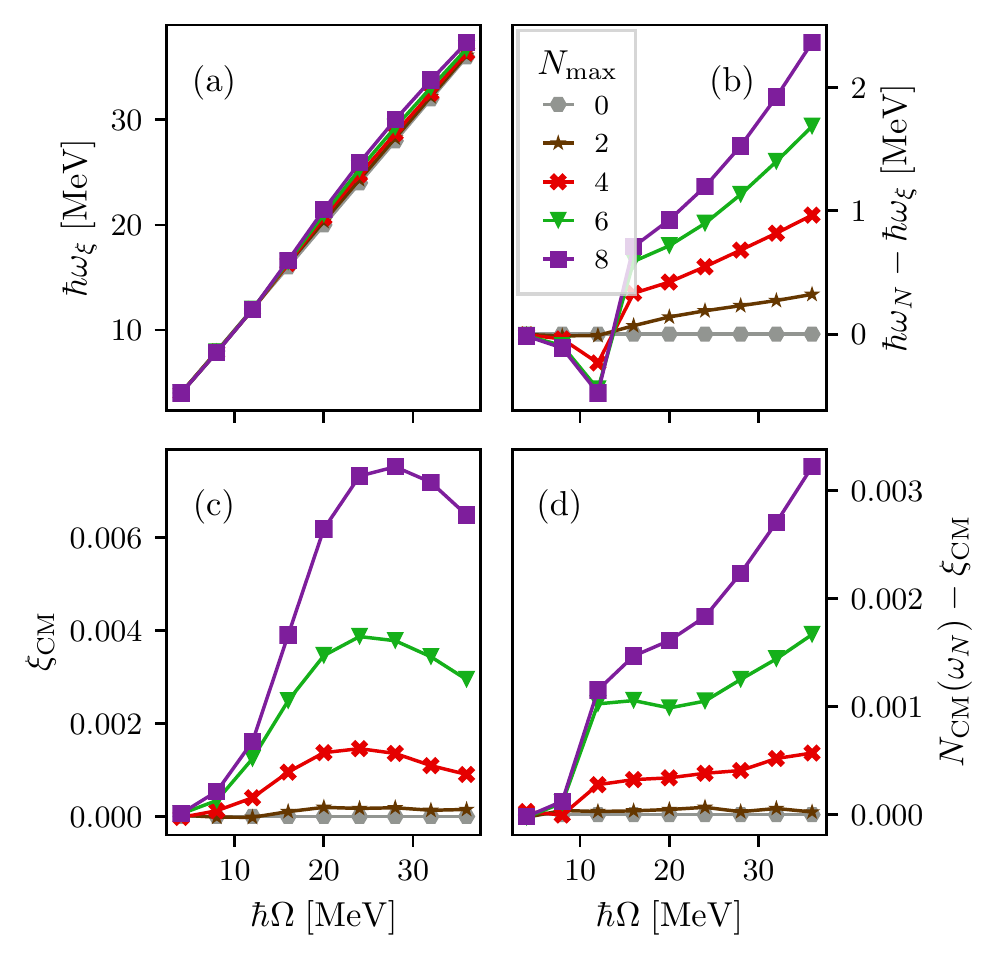}
	\caption{\label{fig:16o_center_of_mass_excitation_combinded}
          \CoM{} analysis for \nuc{16}{O} eigenstates obtained with the \NOTB{} Hamiltonian. The frequency
          $\hbar\wxi$ from Eq.~\eqref{eq:omega_xi} is shown in panel
          (a) and the
          difference $\hbar\wN - \hbar\wxi$ is shown in panel (b). Note that the basis
          frequency $\hw$ is equal to both $\hbar\wN$ and $\hbar\wxi$
          at $\Nmax=0$, as shown
          by the gray line with circle markers.
          The \CoM{} metric \xiCM{} from Eq.~\eqref{eq:XICM} is shown
          in panel (c) and the
          difference $\NCM(\wN) - \xiCM$ in panel (d).
        }
\end{figure}

The evaluated \CoM{} metrics are shown in
Fig.~\ref{fig:16o_center_of_mass_excitation_combinded}, confirming the
satisfactory factorization of the eigenstate. In fact, both \(\NCM{}\)
and \(\xiCM{}\) are orders of magnitude smaller for \nuc{16}{O}
compared to \nuc{4}{He}. Moreover, the HO frequency of the \CoM{}
state is very close to the one for the basis across the frequency
range we explore.
%
%-------------
\subsection{Point proton radii of \nuc{4}{He} and \nuc{16}{O}}
\begin{figure}[htb]
  \includegraphics[width=\columnwidth]{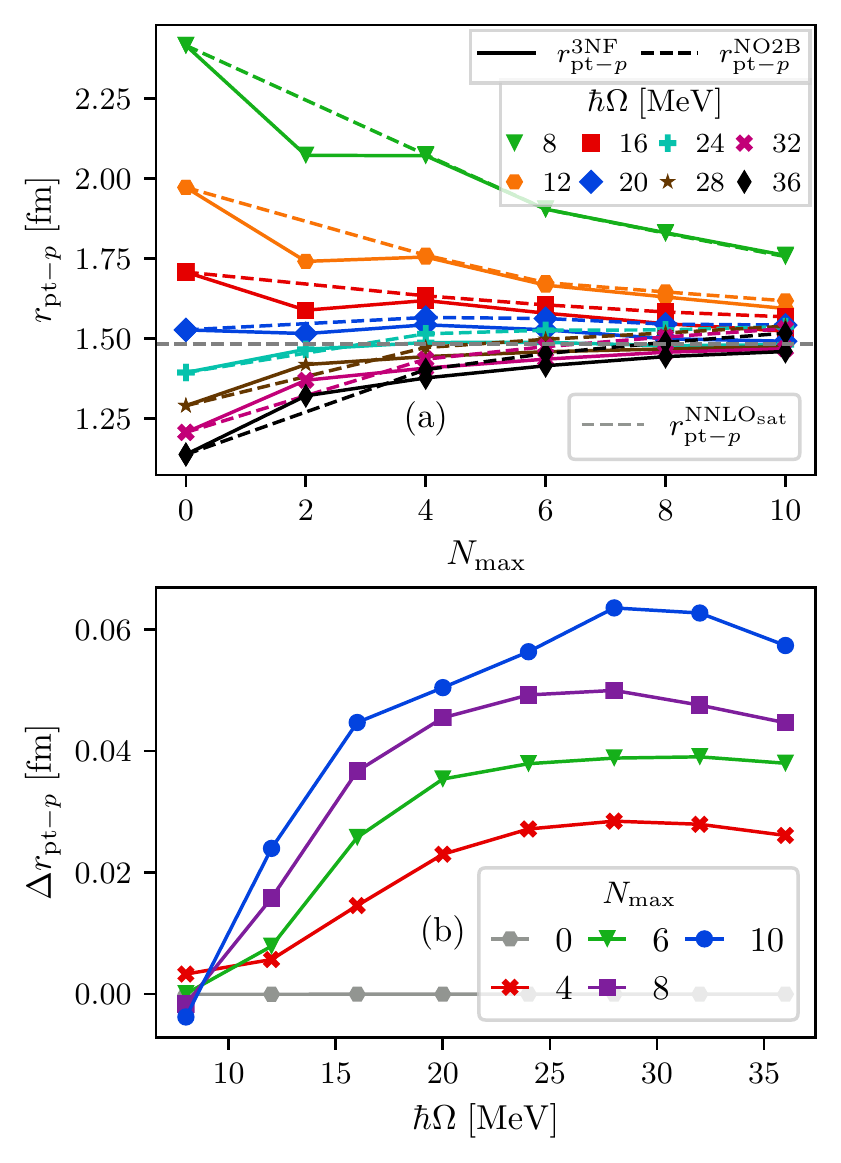}
  \caption{\label{fig:radii_4he}
    The point-proton radius of \nuc{4}{He} computed with either full or
    \NOTB{}-approximated \threeNF{}. 
    The large difference between low and high frequencies
    is due to slower convergence of \NCSM{} calculations at low
    frequencies.
  }
\end{figure}

\begin{figure}[htb]
  \includegraphics[width=\columnwidth]{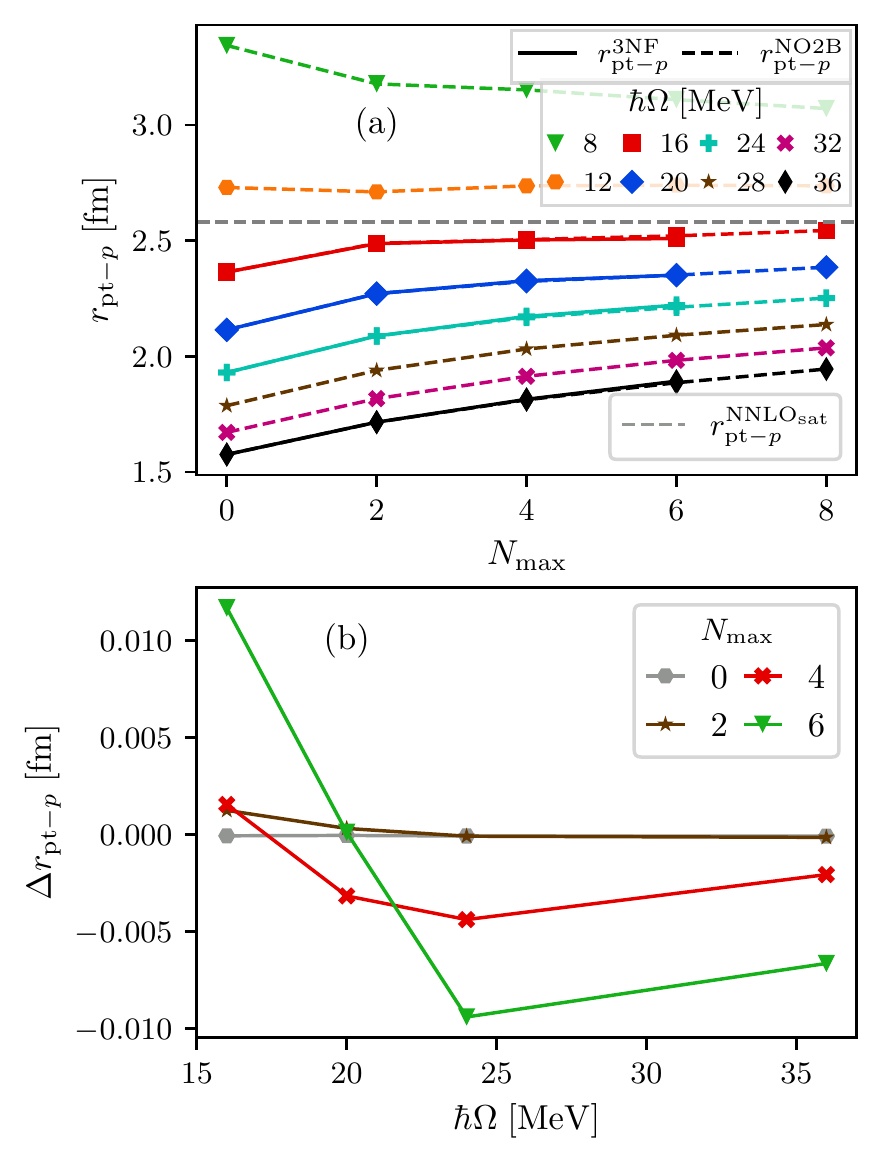}
  \caption{\label{fig:radii_16o}
    The point-proton radius of \nuc{16}{O} computed with either full or
    \NOTB{}-approximated \threeNF{}.
    The large difference between low and high frequencies
    is due to slower convergence of \NCSM{} calculations at low
    frequencies.
  }
\end{figure}

% -------------
%
As a final set of results we also analyze the \NOTB{}-approximation
error in the point-proton radii of \nuc{4}{He} and
\nuc{16}{O}, see Figs.~\ref{fig:radii_4he} and \ref{fig:radii_16o}
respectively.
For \nuc{4}{He} we find a rather large approximation error
and---unlike the results for ground-state energies---there does not
seem to exist an optimal frequency where the error is at a
minimum.

For \nuc{16}{O} we find a good agreement between results with \NOTB{}
approximated and the full Hamiltonian, even though the point-proton
radius shows a much slower convergence rate for very high basis
frequencies. The approximation error is $\lesssim
0.01\;\unit{fm}$. When comparing the radius predictions for
\nuc{4}{He} and \nuc{16}{O} we again find that the \NOTB{}-approximation error
and ensuing \CoM{} contamination decreases with
increasing mass number, as expected. It is also interesting to note
that the maximal \NOTB{}-approximation error for the total energy per
particle and point-proton radius are both reduced by roughly a factor
of 10 when going from $A=4$ to $16$. At least for the range of
oscillator frequencies that we explore here.
%
%-------------
\section{Discussion%
\label{sec:discussion}}
There is a dilemma between the need to include \threeNF{}s in nuclear
calculations to achieve increased physical accuracy and precision, and
the significant increase in computational complexity caused solely by
the inclusion of \threeNF{}s in \emph{ab initio} methods. In this
paper we have studied the \SRNOTB{} approximation of \threeNF{}s, that
aims to reduce the computational complexity to that of \twoNF{}s while
still capturing the most important effects of the \threeNF{}
physics. Our study is performed in the framework of the
\NCSM{} method.

The \SRNOTB{} approximation utilizes Wick's theorem to expand the
\threeNF{} potential in a sum of a constant, one-, two- and three-body
operators that are normal-ordered relative to a non-vacuum reference state
\(\ket{\Psi_{\rm{ref}}}\), taken to be a single \SD{}. In this work
the reference state is constructed in the HO basis and we explore the
sensitivity of computed observables to the choice of the basis frequency.
If the reference state is a good approximation to the ground state of
the nucleus, then the normal-ordered three-body term can be discarded
as it is legitimate to assume that it will have a negligible
contribution to the ground-state energy.

A problem with the \SRNOTB{} approximation is that it breaks the translational
symmetry of the underlying Hamiltonian. In this work we have focused on the
consequences of this symmetry breaking by introducing \CoM{} metrics
and studying the \NOTB{}-approximation error for energies and radii of
\nuc{4}{He} and \nuc{16}{O}.

The main findings and conclusions of this study are:
\begin{itemize}
\item \emph{Translational invariance is explicitly broken in the
    \NOTB{} approximation.}
  The truncation of the normal-ordered Hamiltonian operator introduces
  a \CoM{} dependence of the reference state, which can lead to \CoM{}
  mixing even if a total-energy truncated \NCSM{} basis is used.
\item \emph{Metrics are important for assessing the \CoM{} mixing in
    eigenstates obtained with the \SRNOTB{} approximation.}
  We have found that the previously introduced \(\xi_{\rm{CM}}\) and
  \(N_{\rm{CM}}\) metrics are useful for this purpose.
\item \emph{The comparison of different \CoM{} metrics can reveal more
  information about the details of the \CoM{} factorization.}
$\xiCM = \NCM(\wxi)=0$ imply proper \CoM{}
    factorization with the \CoM{} part in its ground state. However,
    non-zero metrics do not help us determine whether we have a
    mixed state or a linear superposition.
\item \emph{The ability of the \NOTB{} approximation to describe the
  \nuc{4}{He} ground-state energy depends strongly on the
  \NCSM{} basis frequency \hw{}.}
  The \NOTB{}-approximation error is the smallest for \(\hw =
  20\;\unit{MeV}\), but it increases significantly for larger basis
  frequencies. A very weak dependence is observed for smaller
  frequencies. Both \CoM{} metrics indicate negligible \CoM{} mixing
  at small frequencies, and strongly increasing mixing at large ones.
\item \emph{The \CoM{} problem is much less significant for the ground
  state of \nuc{16}{O}.}
  For this system the difference between the \NOTB{}-approximated
  ground-state energies and the full-\threeNF{} ones do not exhibit any
  significant basis-frequency dependency.  Furthermore, there seems to
  be no significant \CoM{} mixing, since both \(\xi_{\rm{CM}}\) and
  \(N_{\rm{CM}}\) are small.
\item \emph{We recommend further investigations of the \CoM{} problem
    in the \NOTB{} approximation also when using other basis functions.}
  In this study we have focused on the \SRNOTB{} approximation with
  a \HO{} basis. However, reference states constructed from other
  single-particle bases might yield better results. In particular the
  Hartree-Fock and the natural orbit bases are being used in some
  many-body
  solvers and results could be analyzed in a similar fashion as in
  this work.
\item \emph{Expectation values of other observables than ground-state
    energies can be strongly affected by the use of the
    \NOTB{}-approximation.}
  Expectation values are computed with respect to the eigenstates, and
  might therefore exhibit a stronger \CoM{}-mixing effect.  While we did compute
  the approximation error for point-proton radii---and found that
  it was particularly large for \nuc{4}{He}---the general effects of
  the \NOTB{} approximation on other observables were not fully
  analyzed in this work.
\end{itemize}
\FloatBarrier
%-------------
\section{Acknowledgment}
We thank S.~R.~Stroberg for useful discussions and
suggestions. We thank P.~Navr\'atil for useful discussions and for
support in the use of the NCSD code.
This work was supported by the Swedish Research Council
  (Grant No. 2017-04234) and the European Research Council (ERC)
  under the European Unions Horizon 2020 research and innovation
  programme (Grant agreement No. 758027). The computations were enabled by
  resources provided by the Swedish National Infrastructure for
  Computing (SNIC) at Chalmers Centre for Computational Science and
  Engineering (C3SE), the National Supercomputer Centre (NSC)
  partially funded by the Swedish Research Council. G.R.J acknowledges
  support by the US Department of Energy under desc0018223 (NUCLEI SciDAC-4
  collaboration).  This research used resources of the Oak Ridge
  Leadership Computing Facility located at Oak Ridge
  National Laboratory, which is supported by the Office of Science of the
  Department of Energy under Contract No. DE-AC05-00OR22725.
%
%-------------
\bibliography{master,temp}

%-------------
\appendix
% -------------
%
% =======
\section{Representations in a truncated basis%
\label{sec:representations}}
\begin{figure}
  \includegraphics[width=\columnwidth]{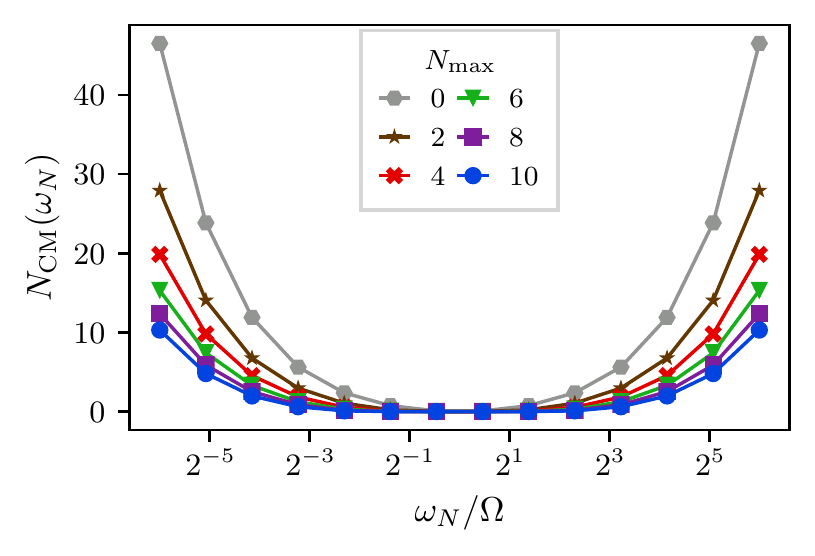}
  \caption{ \label{fig:ho_representation} The smallest eigenvalue
    \(N_{\rm{\CoM}}(\wN{})\), of the \CoM{} \HO{} number operator
    \(\hat{N}_{\rm{\CoM}}(\wN{})\), see Eq.~\eqref{eq:NCM}, computed
    in \NCSM{}-bases with different basis frequencies $\Omega$.
  }
\end{figure}

% ===
%
It is not possible to fully represent a \HO{}-ground state with a frequency
\(\wN{}\) in a truncated \HO{} basis with frequency \(\Omega \neq \wN\).
Therefore, it is possible for the metrics \(\NCM(\wN)\) and \(\xiCM\) to be
non-zero even if the eigenstate is factorized
\begin{equation}
	\ket{\Psi^{\rm{NO2B}}_{\rm{gs}}}_{\rm{NCSM}} =
	\ket{\Psi_{\rm{gs}}}_{\rm{int}}\otimes\ket{\Psi_{\rm{gs}}}_{\rm{CM}}
      \end{equation}
where \(\ket{\Psi_{\rm{gs}}}_{\rm{CM}}\) is a \HO{}-ground state with
frequency \(\wN\).
Here we will study eigenstates of the \HO{} \CoM{}
Hamiltonian~\eqref{eq:HCM}. In particular,
\(N_{\rm{\CoM}}(\wN{})\)---which is the smallest eigenvalue to
\(\hat{N}_{\rm{\CoM}}(\wN{})\) in the current, truncated
\NCSM{} basis---is shown in Fig.~\ref{fig:ho_representation} as a
function of \({\wN{}} / {\Omega}\). Note that the horizontal axis is
logarithmic. It is obvious that this metric is not necessarily zero
even if we have a factorized product state.

We also observe  in
Fig.~\ref{fig:ho_representation} that \(N_{\rm{\CoM}}(\wN{})\) is invariant under the
transformation \(\frac{\wN{}}{\Omega}
\to \frac{\Omega}{\wN{}}\). Here we will demonstrate this algebraically. The
analytical expression for the matrix elements
\(\hat{H}_{\rm{\CoM}}(\wN{})\) in the \CoM{}-coordinate \HO{} basis
\(\ket{\mathcal{N},\mathcal{L}}\) with frequency \(\Omega\) is
\begin{equation}
  \begin{aligned}
    &\bra{\mathcal{N}',\mathcal{L}'}\hat{H}_{\rm{\CoM}}(\wN{})\ket{\mathcal{N},\mathcal{L}} =\\
	  &\quad\bigg[ \left(1+\frac{\wN{}^2}{\Omega^2}\right)
	  \frac{\hw}{2}(2\mathcal{N}+\mathcal{L}+3/2)\delta_{\mathcal{N}',\mathcal{N}}\\
          &\quad-\frac{3}{2}\hbar\wN{}\delta_{\mathcal{N}'\mathcal{N}}
          +\left(1-\frac{\wN{}^2}{\Omega^2}\right)\frac{\hw}{2}\\
      &\quad \times
	  \left(
            \sqrt{\mathcal{N}(\mathcal{N}+\mathcal{L}+1/2)}\delta_{\mathcal{N}'+1,\mathcal{N}}
            \right. \\
      &\qquad \left.
      +\sqrt{\mathcal{N}'(\mathcal{N}'+\mathcal{L}+1/2)}\delta_{\mathcal{N}',\mathcal{N}+1}
      \right) \bigg]
      \delta_{\mathcal{L}',\mathcal{L}}.
  \end{aligned}
\end{equation}
Therefore the matrix elements of \(\hat{N}_{\rm{\CoM}}(\wN{})\) can be written
\begin{equation}
  \begin{aligned}
    &\bra{\mathcal{N}',\mathcal{L}'}\hat{N}_{\rm{\CoM}}(\wN{})\ket{\mathcal{N},\mathcal{L}} =\\
	  &\quad\frac{1}{2}\bigg[ \left(\frac{\Omega}{\wN{}}+
	  \frac{\wN{}}{\Omega}\right)(2\mathcal{N}+\mathcal{L}+3/2)\delta_{\mathcal{N}',\mathcal{N}}
        \\
      &\quad-\frac{3}{2}\delta_{\mathcal{N}'\mathcal{N}}
      +\left(\frac{\Omega}{\wN{}}-\frac{\wN{}}{\Omega}\right) \\
      &\quad \times
	  \left(
            \sqrt{\mathcal{N}(\mathcal{N}+\mathcal{L}+1/2)}\delta_{\mathcal{N}'+1,\mathcal{N}}
            \right.\\
      &\qquad\left.+\sqrt{\mathcal{N}'(\mathcal{N}'+\mathcal{L}+1/2)}\delta_{\mathcal{N}',\mathcal{N}+1}\right)
      \bigg] \delta_{\mathcal{L}',\mathcal{L}}.
  \end{aligned}
\end{equation}
The diagonal is invariant under the transformation
\(\frac{\wN{}}{\Omega} \to \frac{\Omega}{\wN{}}\), but the
off-diagonal terms change sign. However, since the matrix is symmetric
and tridiagonal, the off-diagonal terms will be squared in the
characteristic equation, eliminating the sign change. Thus the
characteristic equation is invariant for
\(\frac{\wN{}}{\Omega} \to \frac{\Omega}{\wN{}}\).  It follows then
that the eigenvalues must be invariant too.
Since \(N_{\rm{\CoM}}(\wN{})\) is the lowest eigenvalue this
demonstrates that it also must be invariant under
\(\frac{\wN{}}{\Omega} \to \frac{\Omega}{\wN{}}\) in accordance with
Fig.~\ref{fig:ho_representation}.

If \(\NCM{}(\wN{})\approx N_{\rm{\CoM}}(\wN{})\)
then \(\ket{\Psi^{\rm{NO2B}}_{\rm{gs}}}_{\rm{NCSM}}\) is separated in a \HO{}-\CoM{}
ground state and some intrinsic state. If, on the other hand,
\(\NCM{}(\wN{}) \gg N_{\rm{\CoM}}(\wN{})\) then
the \CoM{} state is not a \HO{}-ground state.

\end{document}